# Healthcare system resilience and adaptability to pandemic disruptions in the United States

**Lu Zhong**[1,2], **Dimitri Lopez**[1], **Sen Pei**[3], **and Jianxi Gao**[1,2]

[1]Department of Computer Science, Rensselaer Polytechnic Institute, Troy, NY 12180
[2]Network Science and Technology Center, Rensselaer Polytechnic Institute, Troy, NY 12180
[3]Department of Environmental Health Sciences, Columbia University, New York, NY 10032
Email: jianxi.gao@gmail.com

## ABSTRACT

Understanding healthcare system resilience has become paramount, particularly in the wake of the COVID-19 pandemic, which imposed unprecedented burdens on healthcare services and severely impacted public health. Resilience is defined as the system's ability to absorb, recover from, and adapt to disruptions; however, despite extensive studies on this subject, we still lack empirical evidence and mathematical tools to quantify its adaptability (the ability of the system to adjust to and learn from disruptions). By analyzing millions of patients' electronic medical records across US states, we find that the COVID-19 pandemic caused two successive waves of disruptions within the healthcare systems, enabling natural experiment analysis of the adaptive capacity for each system to adapt to past disruptions. We generalize the quantification framework and find that the US healthcare systems exhibit substantial adaptability ($\rho = 0.58$) but only a moderate level of resilience ($r = 0.70$). When considering system responses across racial groups, Black and Hispanic groups were more severely impacted by pandemic disruptions than White and Asian groups. Physician abundance is the key characteristic for determining healthcare system resilience. Our results offer vital guidance in designing resilient and sustainable healthcare systems to prepare for future waves of disruptions akin to COVID-19 pandemics.

## Introduction

Global crises such as climate change, environmental pollution, conflicts, or a global pandemic continue to pose great challenges to the healthcare system[1,2]. These challenges are not solely a result of the escalating scale of these crises but also their prolonged duration with successive disruptions. The COVID-19 pandemic is a convergence of these dual challenges;—an unprecedentedly large-scale crisis that has persisted for over two years, occurred in multiple waves, and resulted in millions of hospitalizations and over 1.2 million deaths in the United States (U.S.) (see Fig.1a). The pandemic has led to disruptions to routine medical services, stemming from factors such as public fear of infections during visits to healthcare facilities[3], stay-at-home policies[4], patient access to care, and limited supply of services. This has led to delays and cancellations in non-COVID-19 emergency services and essential care[4], and the inability to maintain essential services led to adverse and lasting consequences. For instance, in the US, 9.4 million cancer screenings and treatments were either delayed or canceled due to the pandemic[5] and the maternal mortality rate increased from 0.017% in 2019 to 0.032% deaths in 2021[6,7]. Furthermore, the pandemic has disproportionately impacted marginalized groups such as people of color, low-income populations, and those with underlying health conditions[8,9].

To improve overall public health outcomes and mitigate the negative consequences, it is of importance to enhance the resilience of the healthcare system in maintaining essential health services despite disruptions[10–12]. Substantial research has been conducted in this area, including investigations into the conceptual framework of healthcare systems[13–19], examination of the impact of COVID-19 on disrupting health service delivery and degrading health care quality[2,4,20–22], and proposals of measures, such as the social vulnerability index[23,24] and preparedness index[25,26]. However, so far no standard definition and measurement for healthcare system resilience have been established[27].

Resilience is typically defined as the system's ability to absorb and recover from every single disruption[28–31], and also the system's ability to adapt to multiple successive disruptions[27,32] (Table. S1). Existing research on system resilience, spanning various fields including healthcare, ecology, business, and industry, predominantly focuses on the static aspect of absorbing and recovery that enables a system to bounce back after the disruption[33–36]. However, its ability to adapt to disruptions from disaster is rarely explored. As illustrated in Fig.1b, by C.S. Holling et al. in 1970[32,37], adaptability is defined as an adaptive cycle marked by recurrent disruptions and the innate system capability to learn from prior disturbances, ultimately leading to a more resilient system[38]. As recurrent disruptions are rarely observed and recorded, the practical observation and quantitative assessment of this adaptive cycle in real-world systems remains absent. The COVID-19 pandemic, which results in successive waves of disruptions, provides a natural experiment to study the adaptability in a real-world healthcare system.

In this study, we analyze millions of patient records using Electronic Medical Records (EMR) data (see Methods and Table.S2-S3) to create a comprehensive assessment of healthcare systems' resilience and adaptability to disruptions caused by the COVID-19 pandemic in the US. We examine 23 essential health services over a broad range of health needs (see Methods and Table. S4), including chronic disease care (e.g., Alzheimer's disease, Cancer, Heart diseases) and maternal care (e.g., pregnancy). For the healthcare system in each state, we first identify the number of disruptions that the healthcare system encountered and quantify their durations and amplitudes. By comparing a system's performance in absorbing and recovering across disruptions, we are then able to evaluate the system's adaptability and compare it across various health services and distinct patient groups. We also quantified the total system resilience by assessing the loss of patient visits due to the pandemic. Using the COVID-19 pandemic as an example, this study offers a quantification framework for assessing healthcare resilience and adaptability. Table. 1 summarizes the findings and policy implications.

## Results

**Adaptive responses to successive disruptions in healthcare systems.**

The resilience of healthcare systems measures the collective response of diverse entities, such as healthcare providers, hospitals, insurers, pharmacists, the general public, and government entities in sustaining uninterrupted provision of essential services for patients[4,18], as visualized in Fig. 1b. As the COVID-19 pandemic has progressed, it has given rise to several new virus variants, some of which have triggered multiple disruptions in health services. However, the existing quantification framework has primarily focused on a single disruption[35]. Here, we extend the quantification framework to multiple disruptions. Specifically, if there are no external disruptions, the system will maintain its expected performance $P(t)$, represented by non-COVID- 19 patient visits for essential services. As illustrated in the example in Fig. 1c, the actual performance features two disruptions. The dynamics of every single disruption $i$ can be captured by its disruption amplitude $\alpha_i$ (disruption severity), duration $T_i$, disruption rate $u_i$, and recovery rate $v_i$ (see Eq. 1 and Fig. S1). Disruption amplitude and duration gauge the severity of a disruption; higher values mean a more severe impact. Meanwhile, the disruption rate reflects the system's management of the disruption's progress, and the recovery rate indicates how efficiently the system returns to normalcy. Using Fig. 1c as the comparison, a more resilient healthcare system (Fig. 1d) can minimize amplitude and duration, ultimately leading to a smaller loss of patient visits. A higher adaptive healthcare system (Fig. 1e) can slow current disruption rates (or increase recovery rates) than prior ones, as evidenced by $u_2 < u_1$ or $v_2 > v_1$. In Fig. 1f, a healthcare system exhibits both high adaptability and resilience. Please refer to the Methods section for further detailed measurements of the resilience index ($r$) and adaptability index ($\rho$).

Figure 2a shows the trend of patient visits to essential services in the US states. During the pre-pandemic period (from 2017 to 2020), patient visits increased steadily. This increase can be attributed to the increased adoption of certified EMR technology by US hospitals, coupled with the Affordable Care Act's expansion of healthcare resources. We used a predictive model, which incorporates the real-world increasing adoption of EMR technology among physicians within the data, to estimate the expected number of patient visits if the COVID-19 pandemic had not occurred (see Methods, Extended Data Fig. 1, and Fig. S2-S5). In comparison to the expected patient visits $P(t)$, the observed patient visits $O(t)$ started to fluctuate and then decrease sharply at around the beginning of 2020 in Fig. 2a. Among the 49 analyzed states, 40 of them encountered two consecutive disruptions (Extended Data Table. 1). The initial disruption generally occurred between January 2020 and May 2021, followed by a second disruption from June 2021 to the end of 2022. The two disruptions are highly correlated to the waves of the COVID-19 pandemic, with the initial disruption corresponding to the onset of the pandemic, and the second one being exacerbated by the emergence of new, more contagious variants (such as the Omicron variant in the last quarter of 2021). The disruption duration aligns with the results obtained from external datasets on emergency department visits and hospital discharges (Extended Data Fig. 2). All these findings of two disruptions contrast with previous studies that assumed that the COVID-19 pandemic caused a single disruption[2,4,22].

To quantify the system's adaptive response, it first needs to decode each disruption. Fig. 2 and Fig. 3 illustrate the characteristics of the two disruptions. On average, the second disruption tends to have a longer duration and larger amplitude than the initial disruption. However, the disruption rate during the second event was lower than that of the first event. The comparison between these two disruptions suggests that the healthcare system primed during the initial disruption can absorb disruption to decelerate its disruption rates. Regarding recovery, 25 states do not return to expected levels during the first disruption, while 38 of them remain unrecovered until the end of 2022 during the second disruption (Extended Data Table. 1). As the recovery rate is largely unknown, we use the disruption rate to measure the system's adaptability. As depicted in Fig. 2b, systems with a disruption rate smaller than that of the first disruption tend to exhibit higher adaptability. Systems experiencing greater amplitude and extended durations in both disruptions tend to have lower resilience.

Figure 2c-d displays the ranking of adaptability and resilience of healthcare systems in US states. Most states exhibit a positive adaptability index, indicating their ability to improve during the second disruption. Five states standing out with notably negative adaptability indices, suggesting that states didn't have enough resources to better prepare for the following disruptions. Michigan achieves the highest resilience scores with nearly no loss of patient visits ($r = 0.98$), while

Wyoming and Louisiana have the lowest resilience scores ($r = 0.48; r = 0.41$). States that exhibit high adaptivity tend to also demonstrate high resilience (Pearson coefficient= 0.24, p-value: around 0.09) (see Fig. S6). This reflects states with high adaptability being better equipped to handle subsequent disruptions, ultimately resulting in fewer loss of patient visits during the whole period and high resilience. When comparing the indices with the Social Vulnerability Index (SVI)[23,24], we find that the resilience indies are negatively correlated with the SVI for the state (Pearson coefficient= -0.38, p-value: around 0.001) (see Extended Data Fig. 3). These results provide a reasonable validity of the resilience indices. States characterized by higher social vulnerability to disasters tend to have lower resilience scores.

**Adaptability and resilience among essential services**

To explore healthcare system resilience and adaptability in more detail, we group patient visit data based on the services that patients received for their disease. Among essential health services, we focused on chronic disease care and maternal care within the availability of our dataset. For the states with records exceeding a threshold of 1000 patients in both of these services, 82.6% of chronic disease care and 100% of maternal care experienced two disruptions (Extended Data Table. 1). For chronic disease care, 52.2% of states did not achieve recovery by the end of the first disruption, and 83.6% remained unrecovered in the second disruption, which extends until the end of 2022. For maternal care, a more severe situation was observed, with 66.6% of states failing to recover during the initial disruption, and 86.8% of states still had not recovered in the second disruption.

As illustrated in Fig. 3, during the second disruption, healthcare services demonstrated substantial disruption amplitude and prolonged durations. Their enhanced capacity to mitigate disruptions in the second phase results in an overall positive adaptability index for the healthcare system across US states ($\rho = 0.58 \pm 0.05$). Maternal care displayed less adaptability ($\rho = 0.48 \pm 0.10$), in contrast to chronic disease care, which displayed higher adaptability ($\rho = 0.61 \pm 0.06$). For overall resilience, the healthcare system across US states exhibited a moderate level of resilience ($r =$

$0.70 \pm 0.03$). Maternal care thus exhibited lower resilience ($r = 0.74 \pm 0.03$) and experienced disruptions of greater duration, while chronic disease treatment exhibited higher resilience ($r = 0.76 \pm 0.03$). Fig. S4 shows the sensitivity analysis of predictive models across various healthcare services. Fig. S7 shows the resilience and adaptability index of sub-services like COPD, cancer, heart disease, and diabetes and Extended Data Fig. 4 shows the results for services for dialysis.

## Adaptability and resilience among patient groups

Besides maintaining essential services, a resilient healthcare system should be able to provide adequate care to all patients, regardless of their socioeconomic status, race, or ethnicity. We performed sub-group analyses according to race and ethnicity, to measure the number of their visits during the COVID-19 pandemic. More than 22 states experienced two disruptions across all patient groups (Extended Data Table. 1).

Using data from New York state as an example (Fig. 4a), across all patient groups, there is a decrease in disruption rates during the second wave of disruption, demonstrating positive adaptability. Asian population had the lowest adaptability index ($\rho_{Asian} = 0.36 \pm 0.10$) while the white population had the highest ($\rho_{white} = 0.56 \pm 0.06$). This suggests that the Asian population is well-prepared to mitigate the impact of the first disruption, indicating limited room for further improvement and adaptability. Asian population experienced a minimal decline in patient visits and the shortest duration, highlighting the highest resilience score. Meanwhile, the Hispanic and Black populations encountered severe disruptions in amplitude and duration, exhibiting the lowest resilience index. When considering all states (Fig. 4b-e), the Asian population garners the highest resilience score ($r_{Asian} = 0.80 \pm 0.02$), followed by the white population ($r_{white} = 0.74 \pm 0.03$). Conversely, the Black and Hispanic population present the least amount of resilience with scores of $r_{Black} = 0.73 \pm 0.03$ and $r_{Hispanic} = 0.72 \pm 0.03$, respectively. Fig. S4 shows the sensitivity analyses of predictive models across populations according to race.

**Association with pandemic severity, physician shortages, and socioeconomic factors**

The different performance of the healthcare system in US states during the COVID-19 pandemic is complex and multifactorial, with a range of factors, including pandemic severity, healthcare infrastructure, resources, lockdown policies, socioeconomic factors, and the political climate all playing a role. To incorporate these factors into healthcare system planning and decision-making processes, we need to sort out their effects on the resilience index $r$, adaptability index $\rho$, and disruption amplitude $\alpha$. We thus extensively collect state-level COVID-19 cases, physician abundance, and socioeconomic factors to estimate their correlations (Methods). As presented in Table 2, there was a positive correlation between states' resilience index and the abundance of physicians [0.313 (P=0.001)], while there were negative correlations between resilience index and local poverty levels [-0.328 (P=0.001)] and also the unemployment rates [-0.156 (P=0.001)]. We also observed significant correlations between the adaptability index and these factors. For the two pandemic disruptions, their amplitudes exhibit negative associations with physician abundance [-0.241 (P=0.005); -0.205 (P=0.005)] and positive associations with local poverty levels [0.253 (P=0.003); 0.234 (P=0.015)] and uninsurance levels [0.333 (P=0.001); 0.401 (P=0.001)]. The results indicate that states with low physician abundance, high poverty rates, high unemployment, and low insurance coverage are at a higher risk of severe disruption during the pandemic, presenting smaller healthcare system resilience and adaptability.

## Discussion

By analyzing EMR data across US states, we measure the number of patient visits to essential services to analyze the collective response of diverse entities within the healthcare system in the face of consecutive disruptions during the COVID-19 pandemic. Our quantification framework of resilience encompasses key metrics such as the resilience index, adaptability index, and parameters that describe the amplitude and duration of each disruption, as well as the system performance in managing disruption and recovery rate. Our findings reveal that the healthcare system underwent two waves of disruptions, demonstrating an adaptive response wherein lessons from the initial disruption enhance the system's capacity to absorb disruptions and expedite recovery. Over 90% of the healthcare systems performed better during the second disruption. As of the end of 2022, about 77% of health services have yet to rebound to normal levels fully. Consistent with previous research[4], state demographic attributes, such as high poverty levels and high rates of unemployment, were correlated with low resilience and adaptability. The abundance of the physician workforce plays an essential role in determining healthcare resilience and adaptability[39–41]. The findings highlight the importance of strategically organizing the physician workforce during disasters and enhancing collaborations across states[42].

By examining different healthcare services, we found that chronic disease care is more resilient and adaptive than maternal care. This is because chronic disease care needs long-term treatments and is more flexible in delivery across primary care clinics, specialty clinics, and home health services, adding resilience through diverse ways of care[4]. Meanwhile maternal care relies on short-term specialized services, its acute nature makes it more vulnerable to external disruptions[43]. Among the sub-services for chronic disease care (see Fig. S7), services for heart disease management demonstrates lower resilience and adaptability compared to services for conditions such as asthma, COPD, cancer, and diabetes. These findings emphasize the pressing requirement for heightened assistance and focus on heart disease care during a pandemic.

We also examined the healthcare resilience of populations by race and ethnicity. We found that the pandemic severely affected Black and Hispanic populations, leading to harsher disruptions and lower resilience indexes for these populations. These disparities are likely rooted in socioeconomic inequalities that Black and Hispanic communities typically encounter greater obstacles in accessing healthcare services, particularly during external disruptions[44,45]. The findings underscore the importance of enhancing chronic disease care and maternal care to alleviate the exacerbation of inequality, especially for Black and Hispanic communities during crises.

During the COVID-19 pandemic, the number of patient visits to healthcare providers was jointly affected by the supply of services and the patient demand. The availability of resources can limit access to healthcare services, and patient's perception of infection risk may reduce the demand for non-essential services. Our analysis considered the impact of both the supply- and demand-side factors. Using essential service dialysis as the illustration, we disentangle the two factors and offer a rigorous assessment of the availability of delivering services from healthcare infrastructure and resources (the supply side). Patients undergoing dialysis usually need to attend several sessions each week, and their treatment is dependent on consistent and uninterrupted access to these services. Patient visits to dialysis services exhibit two distinct disruptions, differing in duration from other healthcare services. The first disruption occurred from December 2019 to March 2022, while the second disruption commenced in April 2022 with no observed recovery. The dialysis service exhibits delayed disruptions and higher resilience ($r = 0.89$). The results suggest that the healthcare system still underwent two disruptions without significant change in service demand, indicating that the finding of two successive disruptions is replicated.

Nevertheless, the quantification framework has several limitations. First, though the EMR dataset we utilized is extensive within the United States, the incomplete collection and biased data sampling in the dataset can affect the accuracy of our assessment. Specifically, the dataset may primarily represent states that have broadly adopted and implemented the EMR technology, potentially introducing bias into our analysis, particularly in regions where EMR adoption was low. Second, the missing attribute in the dataset can also potentially introduce bias into the analysis. Efforts to mitigate biases in data collection methodologies[46,47] and integrating additional data sources can further validate the findings. Third, our assessment relies on the appropriate selection of a predictive model for forecasting expected patient visits. While we conduct various sensitivity analyses using alternative models, a more sophisticated model is needed to account for uncertainty in data and state variations within the system. Addressing structural uncertainty by considering multiple potential models and assessing their implications for predictions will also be needed for more comprehensive and reliable results.

Several avenues for future research merit exploration for more comprehensive quantification of healthcare system resilience. It is crucial to consider excessive recovery as a novel dimension of system resilience, aiming not only to return to its original state but to recover lost visits and progress towards a more optimal state (Fig. S8). It is necessary to untangle the intertwined impacts of successive disruptions, particularly when disruptions stem from different crises. Furthermore, there is also a necessity to broaden the scope beyond COVID-19 waves to comprehend healthcare system performance across different crises, especially for future guiding healthcare system resilience against more deadly diseases associated with climate change. For example, conducting a comparative analysis of responses to the 2003 SARS outbreak and the 2019 COVID-19 pandemic could serve as a guide. Additionally, while our measure reflects the overall resilience of the healthcare system; it is crucial to exclude demand factors to concentrate solely on the dynamics of service availability when evaluating the resilience for specific hospitals and other healthcare facilities. Moreover, our macroscopic measures are limited in offering higher-resolution guidance on each system component. Future studies should delve into sophisticated computational methods and consider additional factors such as the intricate connections between healthcare facilities, resource management, the healthcare workforce, supply chain dynamics, and the quality of patient care[18,41]. Lastly, our measures can only be assessed retrospectively. Future endeavors should strive to provide real-time estimates for proactive decision-making.

In summary, our study provides a quantification framework for healthcare system resilience that can be applied generically to various healthcare services, patient groups and different regions[48,49]. The proposed Adaptability Index and Resilience Index allow characterization of healthcare system performance during and across multiple disruptions. These two indices can also offer a valuable complement to existing crisis management tools, such as Social Vulnerability Index (SVI)[24] and Preparedness Index (PPI)[25,26], which predominantly focus on either the socioeconomic status of a region or the availability of healthcare resources before disruption strikes. In brief, this framework could provide policymakers with the essential insights to make informed adaptation to successive disruptions caused by prolonged disasters[17,50,51].

# Acknowledgements

The data, technology, and services used in generating these research findings were generously supplied pro bono by the COVID-19 Research Database partners, who are acknowledged at https://covid19researchdatabase.org/. This work was supported, in whole or in part, by the Bill & Melinda Gates Foundation (No. CORONAVIRUSHUB-D-21-00120). Under the grant conditions of the Foundation, a Creative Commons Attribution 4.0 Generic License has already been assigned to the Author Accepted Manuscript version that might arise from this submission. We also acknowledge the support of the USA National Science Foundation (No. 2047488) and the Rensselaer- IBM AI Research Collaboration.

# Author Contributions Statement

L.Z., D.L., and J.G. conceived the project and designed the study. D.L. and L.Z. performed the data analyses. L.Z. and D.L. wrote the first draft of the manuscript. L.Z., S.P., and J.G. contributed to interpreting the results and improving the manuscript. J.G. was the lead writer of the manuscript.

# Competing Interests Statement

The authors declare no competing interests.

**Table 1.** Policy Summary

| | |
|---|---|
| Background | Resilience is defined as the system's ability to absorb and recover from each single disruption, as well as its adaptability—to adapt or transform itself to better respond to multiple successive disruptions. However, despite the COVID-19 pandemic bringing multiple waves of disruptions to healthcare systems, there has been limited research dedicated to quantifying their adaptability to successive disruptions and their resilience during the pandemic. |
| Main findings and limitations | By analyzing extensive Electronic Medical Record (EMR) data across US states, we find that the COVID-19 pandemic led to two successive disruptions within healthcare systems. We generalized the quantification framework and assessed the resilience of healthcare systems across various states for different essential services and populations according to race and ethnicity. The results show that healthcare systems demonstrate significant adaptability but only a moderate level of resilience. Services for chronic disease treatment exhibit higher resilience compared to maternal services. Black and Hispanic populations were most affected by severe disruptions when compared to White and Asian groups. By examining the relationship between system resilience and factors such as pandemic severity, physician shortages, and socioeconomic variables, we identified physician abundance as the pivotal characteristic influencing healthcare system responses. Limitations of this study include the bias in data collection process and missing attributes within the EMR dataset, potentially compromising the accuracy and reliability of predictive trends related to patient visits based on historical data. |
| Policy implications | Our results highlight the importance of improving the system's adaptability to effectively respond to ongoing disruptions, especially for maternal care, minority populations in the US, and states with a scarcity of physicians, high poverty, and low employment. The Resilience and Adaptability Indexes we have introduced, founded on a dynamic perspective, complement existing metrics like the Social Vulnerability Index (SVI) and Preparedness Index, offering guidance for future disasters that spikes waves of disruptions akin to COVID-19 pandemics. |

**Table 2.** Pearson correlation coefficients assessing the relationships between system adaptability/resilience and pandemic severity, physician shortages, and socioeconomic factors in U.S. states. Significant correlations, determined by a two-sided test and indicated by a P-value less than the threshold of 0.05, are highlighted.

|  | COVID-19 cases | Physician per 100,000 | Poverty percentile | Unemployment percentile | Uninsurance percentile | Age≥65 percentile | Age≤17 percentile | Minority percentile |
|---|---|---|---|---|---|---|---|---|
| Adaptability index | -0.033 (p=0.507) | **0.102 (p=0.039)** | **-0.082 (p=0.046)** | **-0.146 (p=0.003)** | -0.010 (p=0.828) | 0.040 (p=0.404) | -0.009 (p=0.847) | -0.0046 (p=0.350) |
| Resilience index | -0.054 (p=0.278) | **0.313 (p=0.001)** | **-0.328 (p=0.001)** | **-0.156 (p=0.001)** | **-0.47 (p=0.001)** | **0.116 (p=0.019)** | **-0.209 (p=0.001)** | **-0.217 (p=0.009)** |
| Amplitude $\alpha$ (1st disruption) | 0.114 (p=0.198) | **-0.241 (p=0.005)** | **0.253 (p=0.003)** | 0.093 (p=0.295) | **0.333 (p=0.001)** | -0.116 (p=0.191) | **0.175 (p=0.047)** | 0.054 (p=0.369) |
| Amplitude $\alpha$ (2nd disruption) | 0.132 (p=0.178) | **-0.205 (p=0.035)** | **0.234 (p=0.015)** | **0.206 (p=0.034)** | **0.401 (p=0.001)** | -0.042 (p=0.668) | 0.143 (p=0.144) | 0.103 (p=0.292)[1] |

---

[1] Significant correlations are highlighted in bold.

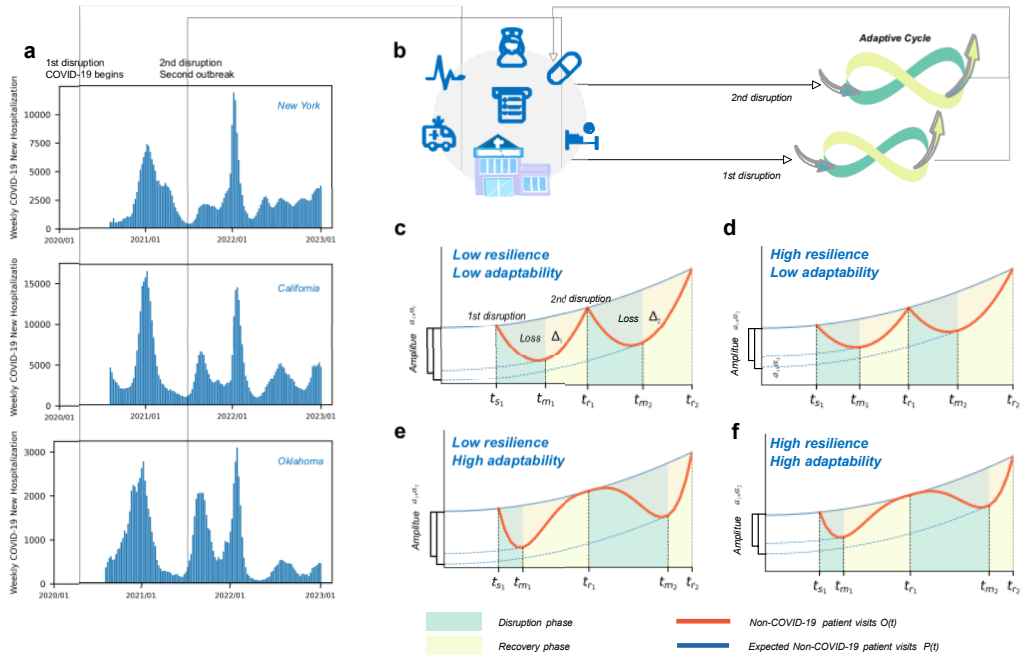

**Figure 1. Adaptive responses to successive disruptions in healthcare systems.** As seen in the data on COVID-19 hospitalizations (**a**), the pandemic brought successive disruptions for healthcare systems. (**b**) The adaptive response cycle within the healthcare system involves multiple cycles of disruption and recovery, leading to enhanced resilience through learning from prior disruptions. By examining the system's performance in maintaining non-COVID-19 patient visits (**c,d,e,f**), we assess its adaptability and resilience during successive disruptions. With two disruptions as the example in (**c**), when the disruption event $i$ (e.g., onset of COVID-19 pandemic and second wave of the outbreak) occurs at time $t_{s_i}$, the healthcare system's performance (non-COVID-19 visits) begins to decline, reaches a negative peak and then returns to the target performance level at $t_{r_i}$. Each disruption $i$ undergoes a disruption and recovery phase, characterized by disruption amplitude ($\alpha_i$), disruption duration ($T_i = t_{r_i} - t_{s_i}$), disruption rate ($u_i$), and recovery rate ($v_i$). (**c**) A system with low resilience and low adaptability. (**d**) A system with high resilience is characterized by smaller amplitude/duration. (**e**) A system with high adaptability is characterized by its ability to better absorb disruptions $u_{i+1} < u_i$ or quicker recover $v_{i+1} > v_i$ during the second disruption. (**f**) A system with higher resilience and adaptability.

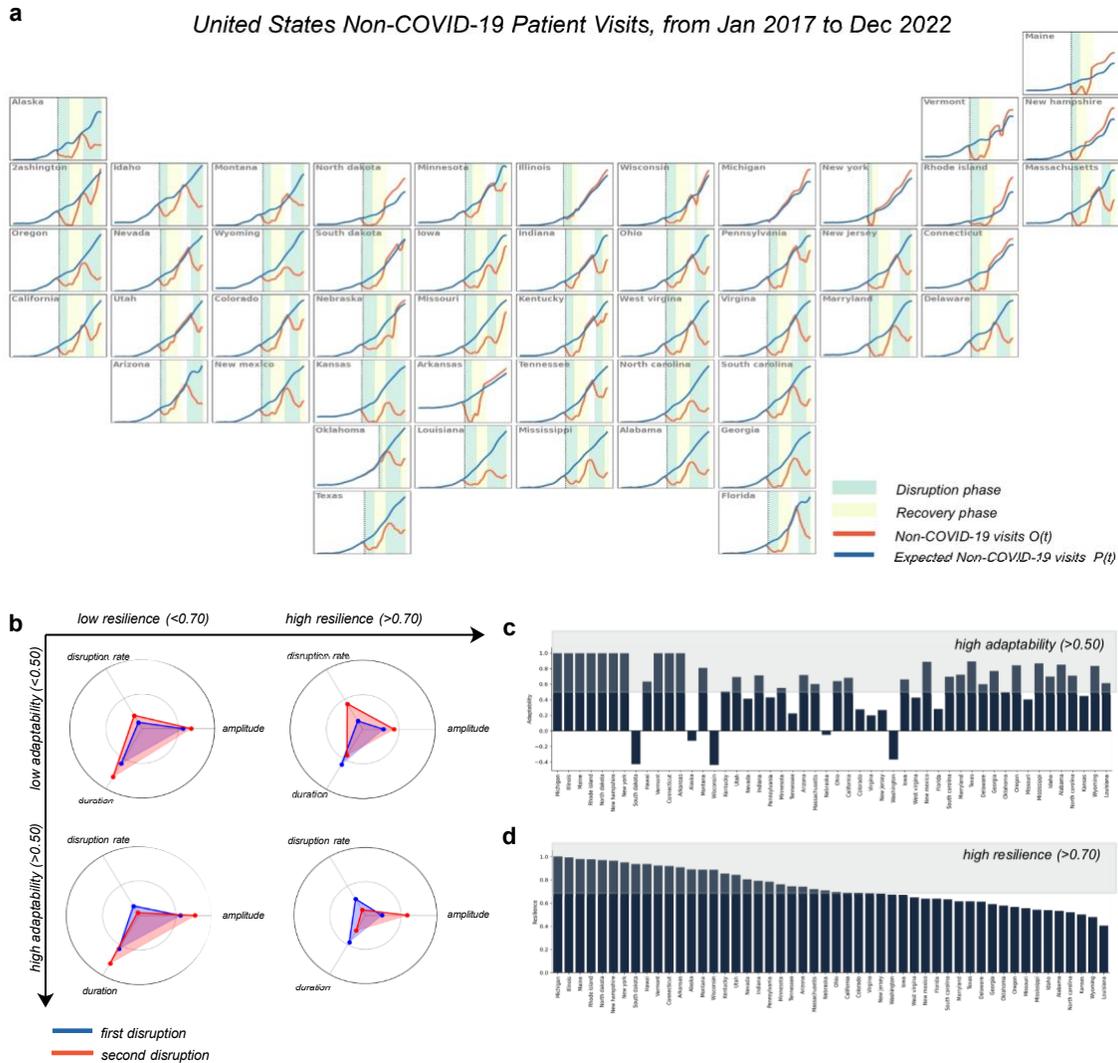

**Figure 2. Adaptability and resilience assessment of US healthcare systems.** (**a**) Temporal trend of non-COVID-19 patient visits across US States, January 2017 to December 2022. Among 49 analyzed states, 40 (81.63%) of them experience two successive disruptions. The initial disruption generally occurred between January 2020 and May 2021, followed by a second disruption starting from June 2021, and 38 of them were not recovered by the end of 2022. (**b**) Disruption amplitude $\alpha_i$, duration $T_i$, and disruption rate $u_i$ characterize each disruption for states in high or low levels of adaptability and resilience. (**c**) State rankings for adaptability with index $\rho$ in the range of $[-1, 1]$, where a positive value indicates a capacity to adapt. (**d**) State rankings for resilience with index $r \in [0,1]$, where a higher value indicates a greater capacity to encounter disruptions and sustain the volume of visits. For better comparison, the plots in (**c,d**) are in descending order of resilience index. A system is considered to have high adaptability when $\rho > 0.5$ and high resilience when $r > 0.7$, highlighted in a gray band.

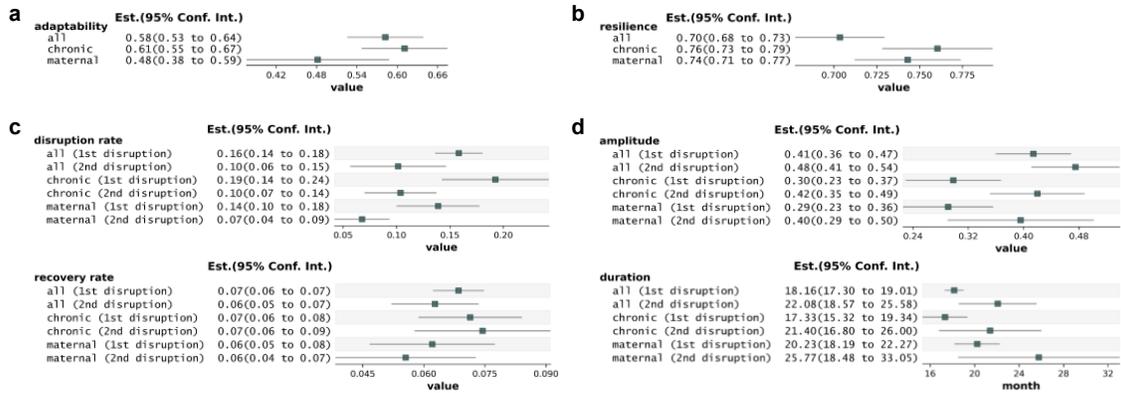

**Figure 3. Adaptability and resilience assessment for essential services.** To measure healthcare system responses for essential services, we categorize non-COVID-19 patient visits into two specific services across US states: chronic disease care and maternal care. The 'all' category encompasses all services. **(a)** The average adaptability ($\rho$). **(b)** The average resilience ($r$). **(c,d)** Parameters characterize each disruption for services. The parameters include disruption amplitude ($\alpha_i$), disruption duration ($T_i$), disruption rate ($u_i$), and recovery rate ($v_i$). The adaptability and resilience indices and parameters across all states are expressed by the average, along with the lower and upper limits of the 95% confidence level.

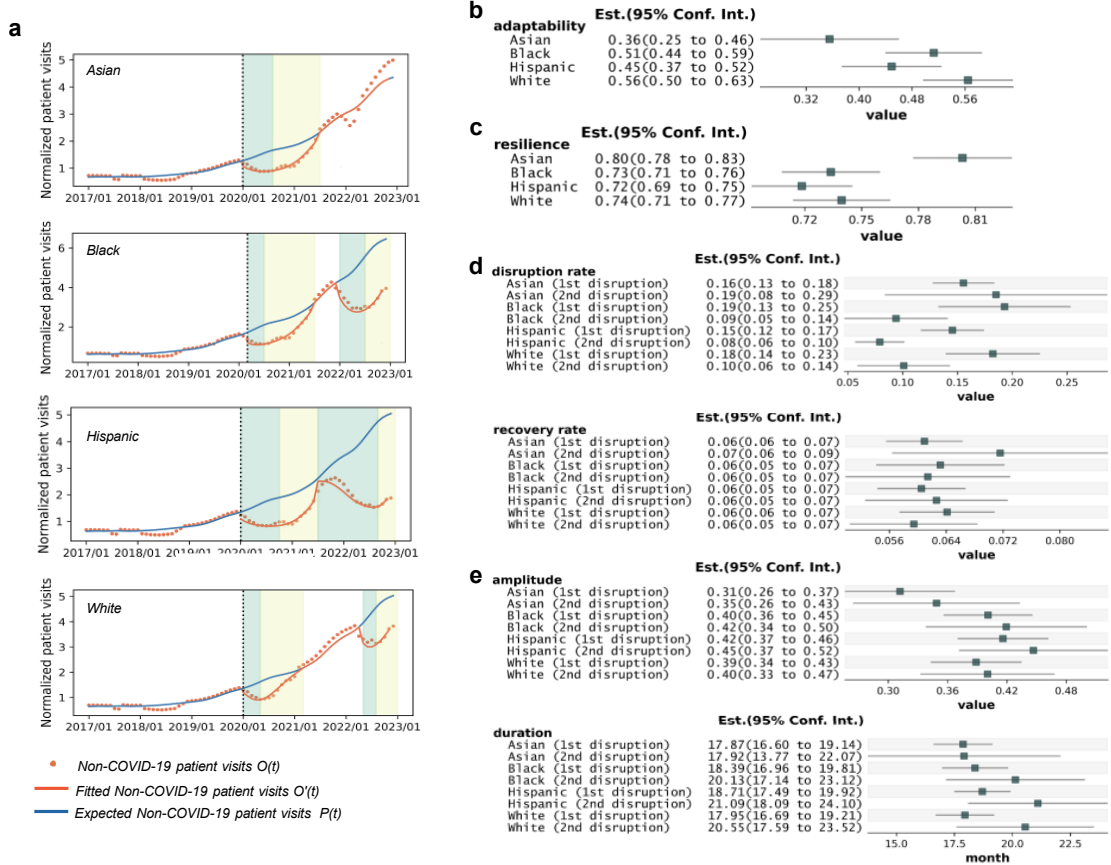

**Figure 4. Adaptability and resilience assessment for patient groups in different races and ethnicities.** To measure healthcare system responses for patients in different races and ethnicities, we group by non-COVID-19 visits according to patients' attributes across US states. **(a)** Illustration of the temporal trend of non-COVID-19 patient visits for Asian, Black, Hispanic, and White groups in New York state. **(b)** The average adaptability ($\rho$). **(c)** The average resilience ($r$) of patient groups. **(d,e)** The parameters characterize each disruption. The parameters include disruption amplitude ($\alpha_i$), disruption duration ($T_i$), disruption rate ($u_i$), and recovery rate ($v_i$). The adaptability and resilience indices and parameters across all states are expressed by the average, along with the lower and upper limits of the 95% confidence level.

## Methods

**Data.** The EMR dataset we utilize, referred to as the Healthjump dataset, is provided by the COVID-19 Research Database[52]. Healthjump is a data integration platform exporting electronic medical records and practice management systems. It aggregates data from over 70,000 hospitals and clinics, as well as more than 1,500 healthcare organizations, covering all states in the US. With records dating back to 1995, the dataset encompasses information on over 33 million patients and 1 million healthcare providers. It includes a wide array of patient data, including diagnoses, procedures, encounters, and medical histories sourced from participating members of the Healthjump network. Additionally, it contains patients' socio- demographic attributes, such as race, gender, and others. The dataset details, statistics, and limitations are presented in Table. S2 and S3.

In examining the correlation between states' resilience index with COVID-19 infections, physician abundance, and socio-demographic factors, we first gather cumulative infection cases in each state from the Johns Hopkins Coronavirus Resource Center[53]. We collect physician abundance data regarding physician numbers for each state, drawing from the 2019 State Physician Workforce Data[54]. The social factors analyzed include poverty levels, unemployment rates, uninsured levels, the proportion of the youth (less than 17), the proportion of the elderly (greater than 65), and the proportion of minority populations are all sourced from the CDC/ATSDR Social Vulnerability Index[24].

**Ethics statement.** Ethical approval was not required for this study as the data used for analysis from the fully anonymized Healthjump EMR database[52]. The database complies with Health Insurance Portability and Accountability Act (HIPAA) of 1996, ensuring the protection of patient information through strict Privacy Policy and agreements with patient and healthcare provides. Therefore, the ethical approval was not needed, as the database has no identifiable information of individual patients.

**Demographic information.** Demographic information was restricted to race and ethnicity. For the EMR-reported race and ethnicity, patients were designated by themselves or by a healthcare provider. For the race, patients were designated as "American Indiana or Alaska Native", "Asian", "Black or African American", "White", "Native Hawaiian or other Pacific Islander", and "other

Race". For the ethnicity, patients were designated by "Hispanic or Latino", "Not Hispanic or Latino", and "Unknown". For simplification, we consider the main race/ethnic categories in the study.

**Essential health services of concern.** Essential health services encompass vital healthcare provisions crucial for enhancing and preserving public health[1,4]. These services typically comprise maternal care, chronic disease management, diagnostic and laboratory services, vaccination, primary care, emergency care, and others. Due to limitations in data availability and the exclusion of data related to care for COVID-19 patients, our study focuses solely on chronic disease care and maternal care across a spectrum of 23 diseases (refer to Table. S4). We analyzed the absolute number of visits for these services. Additionally, we normalized the data to mitigate the impact of state population size and policies related to joining Healthjump's EMR system in our analyses.

**Quantification framework for successive disruptions.** We employ mathematical models that provide key parameters to characterize the system's behavior during disruption and recovery processes. While numerous models exist[55], the beta family equations uniquely offer flexibility (see Fig. S1). Building upon the framework that describes the system's behavior under a single disruption[35], we generalize this framework to multiple successive disruptions,

$$O(t) = \begin{cases} P(t); & 0 \leq t < t_{s_1} \\ P(t) - \alpha_1 \frac{(\theta_1+\vartheta_1)^{\theta_1+\vartheta_1}}{\theta_1^{\theta_1}+\vartheta_1^{\vartheta_1}} \left(\frac{t}{T_1}\right)^{\theta_1} \left(1-\frac{t}{T_1}\right)^{\vartheta_1}; & t_{s_1} \leq t < t_{r_1} \\ P(t) - \alpha_2 \frac{(\theta_2+\vartheta_2)^{\theta_2+\vartheta_2}}{\theta_2^{\theta_2}+\vartheta_2^{\vartheta_2}} \left(\frac{t}{T_2}\right)^{\theta_2} \left(1-\frac{t}{T_2}\right)^{\vartheta_2}; & t_{s_2} \leq t < t_{r_2} \\ \cdots \cdots \\ P(t); & t > t_{r_n} \end{cases} \quad (1)$$

where $O(t)$ is the actual observed performance of patient visits and $P(t)$ is the predicted performance if the pandemic didn't occur. Suppose there are *n* disruptions. Each disruption *i* has an amplitude $\alpha_i$, which is the scale factor, defined as the severity of disruption on the system. The other two parameters $\theta_i$, and $\vartheta_i$ determine the curve's shape for disruption and recovery respectively, as reflected in the system's ability to manage the processes. Within the same duration $T_i$ ($T_i = t_{r_i} - t_{s_i}$), if $\theta_i = \vartheta_i$, a symmetric disruption and recovery occur; if $\theta_i > \vartheta_i$, a slow disruption is followed by a fast recovery; if $\theta_i < \vartheta_i$, a fast disruption is followed by a slow recovery. We define disruption rate as $u_i = \frac{1}{\theta_i T_i}$ and its recovery rate as $v_i = \frac{1}{\vartheta_i T_i}$. A smaller value of $\theta_i$ or $\vartheta_i$ leads to quick disruption or recover. A shorter duration $T_i$ results in a rapid deterioration of performance to its maximum extent.

To assess the system's performance across $n$ disruptions, we introduce the adaptability index in terms of disruption rate across consecutive disruption $i$ and $i + 1$,

$$\rho = \frac{1}{n}\sum_{i=1}^{n} \frac{-(u_{i+1}-u_i)}{\max(u_{i+1},u_i)} \tag{2}$$

where $\rho \in [-1,1]$. Specifically, $\rho > 0$ indicates that the system exhibits adaptability with $u_{i+1} < u_i$, signifying that the rate of disruption $i + 1$ is smaller than that of disruption $i$. Conversely, $\rho < 0$ suggests that the system lacks adaptability with $u_{i+1} > u_i$, indicating that the rate of disruption $i + 1$ is larger than that of disruption $i$. Higher values of $\rho$ signify an increased level of adaptability of the system. In the case of a single disruption, $\rho = 1$. The adaptability index can also be measured in terms of recovery rate. As the recovery rate is largely unknown in our results, we only use the disruption rate to measure the system adaptivity.

Following the classic way[29,30,55,] we measure system resilience in terms of performance loss as

$$r = 1 - \frac{\int_{t_s}^{t_r}[P(t)-O(t)]dt}{\int_{t_s}^{t_r} P(t)dt} \tag{3}$$

where $r \in [0,1]$. The term $\int_{t_s}^{t_r}[P(t) - O(t)]dt$ is the total loss between expected $P(t)$ and observed patient visits $O(t)$. The division by integral of expected performance $\int_{t_s}^{t_r} P(t)dt$ normalizes the results to a range between 0 and 1. A value of 1 indicates no performance loss (perfect resilience), while 0 indicates a complete loss.

**Predictive model for quantification framework.** For the analysis of healthcare resilience during the COVID-19 pandemic, we extracted data from 2017 to the end of 2022. Commencing from January 2020, the onset of the pandemic, we designate this as the starting date, denoted as $t_s$, The dataset is subsequently divided into two distinct periods: the pre-pandemic period ($t < t_s$, from January 2017 to December 2019) and the pandemic period ($t \geq t_s$, from January 2020). With the growing adoption of EMR technology in U.S. hospitals and increased accessibility to healthcare facilities due to the Affordable Care Act, there has been a noticeable rise in patient visit volumes in our datasets. To capture the expected patient visits in Eq. (1), we leverage the number of physicians in the data that participate in EMR technology, state population, and physician-to-population ratio[54] to assess expected patient visits,

$$P(t) = \mathcal{P}_v(t)\sigma N \qquad (4)$$

where $\mathcal{P}_v(t)$ is the monthly number of physicians that adopted EMR technology in the state, $N$ is the state population, and $\sigma$ is the state Physician-to-Population Ratio, adjusted by monthly visits frequency in the pre-pandemic period. Through the models, we can predict the expected patient visits that digital health platforms would accumulate if no disruptions occurred beyond $t_s$. We also employ the innovation (EMR) adoption model[56] and time-series model[57], i.e., the generalized logistic model and exponential smoothing model, and consider the seasonality for the sensitivity test. The comparisons of predictive models are provided in Extended Data Fig. 1 and Fig. S2-S5. To smooth out seasonal fluctuations and identify underlying trends or patterns, we use the three-month moving average on data. For comparison across states and services, the volume of patient visits at $t = 0$ is normalized to 1.

**Identification of disruptions.** To fit the observed performance $O(t)$ with Eq. (1), we first identify the number of disruptions $n$ and the duration $T_i$ of each disruption $i$. The disruptions are identified through segmented least squares, which enables the detection of both the peak of decline and subsequent increase in performance and divides the duration $T_i$ into disruption and recovery phases. Disruptions failing to meet the criteria will be excluded: (1) those lasting less than three months (where $T_i < 3$) and (2) those with a minimum performance loss ratio below 5%. These two criteria are used to exclude disruptions with non-significant losses and those attributed to data fluctuations caused by seasonal effects (Check Fig. S5 for the sensitivity test of criteria). Then we infer the parameters $\alpha_i$, $\theta_i$ and $\vartheta_i$ for the observed performance for each disruption $i$ by using the iterative optimization method[58,59].

**Statistical analysis.** To demonstrate the resilience and adaptability of various services and races, we represent their average across analyzed states with the lower and upper limits of 95% confidence level in Fig. 3 and Fig. 4. For a total of $M$ analyzed state, the average of resilience for services is calculated as $r^i = \frac{\sum_m r_m^i}{M}$, where $r_m^i$ represents the resilience at state $m$ for services $i$. Similarly, the average of resilience for racial groups is calculated as $r^j = \frac{\sum_m r_m^j}{M}$, where $r_m^j$ represents the resilience at state $m$ for race $j$.

We employ the Pearson correlation coefficient to gauge the correlation between state resilience (or adaptability) with various factors, such as COVID-19 infections, physician abundance, socio-demographic factors, and SVI. This coefficient, ranging between –1 and 1, quantifies both the strength and direction of the relationship between two variables.

## Data availability

The electronic medical record dataset that supports the findings of this study is available from the Healthjump database, provided by the COVID-19 Research Database consortium (https://covid19researchdatabase.org/). However, restrictions apply accessing these data, which were used under license for the current study. The EMR dataset is not publicly available. The data on COVID-19 infection cases in each state is collected from the Johns Hopkins Coronavirus Resource Center (https://github.com/CSSEGISandData/COVID-19). The general physician abundance

**28/31**

data regarding physician numbers in each state is collected from the 2019 State Physician Workforce Data (https://www.aamc.org/data-reports/workforce/report/state-physician-workforce-data-report). The socio-demographic factors in each state are collected from the CDC/ATSDR Social Vulnerability Index (https://www.atsdr.cdc.gov/placeandhealth/svi/index.html). For validation, external summary datasets on patient visits to physicians, emergency departments, and the number of hospital discharges during the pandemic are sourced from the National Center for Health Statistics (https://www.cdc.gov/nchs/index.htm) and the US Census Bureau (https://www.census.gov/). For the results dashboard, please see website ResilienceHealthSys.com.

## Code availability

The code used in the study for the quantification framework is provided at https://github.com/lucinezhong/healthcare_resilience_quantification_framework.